\documentclass[12pt, oneside, reqno]{amsart}
 \usepackage{graphicx}
\usepackage{amsmath,amsthm,amsfonts,amscd,amssymb,mathrsfs,hyperref}
\usepackage[all]{xypic}
\usepackage{amscd}   
\usepackage[all]{xy} 
\usepackage{booktabs} 
\usepackage{physics}
\usepackage{hyperref}
\usepackage{listings}
\usepackage{algorithm2e}
  \hypersetup{colorlinks=true,citecolor=blue, urlcolor= cyan}
\usepackage[foot]{amsaddr}

\usepackage{ytableau}
\usepackage[enableskew]{youngtab} 

\usepackage{nicefrac}
\usepackage{xfrac}

\usepackage{color}

\makeatletter
\newtheorem{rep@theorem}{\rep@title}
\newcommand{\newreptheorem}[2]{%
\newenvironment{rep#1}[1]{%
 \def\rep@title{#2 \ref{##1}}%
 \begin{rep@theorem}}%
 {\end{rep@theorem}}}
\makeatother

\newtheorem{theorem}{Theorem}[section]
\newreptheorem{theorem}{Theorem}
\newreptheorem{lemma}{Lemma}

\newtheorem{lemma}[theorem]{Lemma}

\newtheorem{proposition}[theorem]{Proposition}

\newcommand{\PP}{\mathbb{P}}

\newcommand{\CC}{\mathbb{C}}

\def\phi{ \varphi }

\theoremstyle{definition}
\newtheorem{definition}[theorem]{Definition}

\theoremstyle{remark}
\newtheorem{remark}[theorem]{Remark}

\newcommand{\ccirc}[1]{\xymatrix@1{+<1ex>[o][F-]{#1}}}

\title{Entanglement and non-locality of four-qubit connected hypergraph states}
\author{Gr\^ace Amouzou ($\star$,$\diamond$)}
\author{Jeoffrey Boffelli ($\square$)}
\author{Hamza Jaffali ($\triangle$)} 
\author{Kossi Atchonouglo ($\diamond$)}
\author{Fr\'ed\'eric Holweck ($\star$)}
\address[$\star$]{Laboratoire Interdisciplinaire Carnot de Bourgogne, ICB/UTBM, UMR 6303 CNRS,
Universit\'e Bourgogne Franche-Comt\'e, 90010 Belfort Cedex, France
}
\address[$\diamond$]{Laboratoire de Mod\'elisations Math\'ematiques et Applications, Universit\'e de Lom\'e, Togo}
\address[$\square$]{Laboratoire Ondes et Milieux Complexes, CNRS–Universit\'e Le Havre Normandie, Le Havre, France} 
\address[$\triangle$]{Femto-st, UMR 6174 CNRS, Universit\'e Bourgogne Franche-Comt\'e, 90010 Belfort Cedex, France}
\email{frederic.holweck@utbm.fr}
\begin{document}

\maketitle

\begin{abstract}
We study entanglement and non-locality of connected four-qubit hypergraph states. One obtains the SLOCC classification from the known LU-orbits. We then consider Mermin's  polynomials and show that all four-qubit hypergraph states exhibit non-local behavior. Finally, we implement some of the corresponding inequalities on the IBM Quantum Experience.
\end{abstract}

\section{Introduction}
It is known since the work of Verstraete et al. \cite{verstraete} that the Hilbert space of the four-qubit states has an infinite number of orbits under Stochastic Local Operations and Classical Communictation  (SLOCC) that can be described by 9 families -- 6 of them depending on parameters. The most generic four-qubit quantum states being the $G_{abcd}$ class which depends on four-parameters:
\begin{equation}
 \resizebox{.9\hsize}{!}{$\ket{G_{abcd}}=\frac{a+d}{2}(\ket{0000}+\ket{1111})+\frac{a-d}{2}(\ket{0011}+\ket{1100})+\frac{b+c}{2}(\ket{0101}+\ket{1010})+\frac{b-c}{2}(\ket{0110}+\ket{1001}).$}
\end{equation}
Since then the four-qubit classification has generated a large amount of work including alternative perspectives on the classification itself \cite{miyake_hyperdet,chterental2006normal,li2007slocc,lamata2007inductive,chen2013four}, invariants-covariants approches of the classification \cite{luque2003polynomial,viehmann2011polynomial,dhokovic2009polynomial}, geometric intepretations \cite{levay2006geometry,holweck2014entanglement,holweck2017entanglement,miyake_hyperdet} and connection with others domains in physics \cite{borsten2010four,borsten2012black,belhaj2018four}.
Hypergraph states are quantum states that generalize the notion of graph states where qubits composing the system are given by vertices and the interaction between the qubits/vertices are described by edges or hyperedges (See Sec. \ref{hypergraph} for the definition). Quantum hypergraph states have been introduced in \cite{rossi2013quantum} and, like graph states, are nowadays recognized as a resource for Measured Based Quantum Computation (MBQC) \cite{takeuchi2019quantum}. Properties of hypergraph states in terms of entanglement and non-locality have been investigated in \cite{guhne2014entanglement,gachechiladze2016extreme}. An exhaustive reference for the study of hypergraph states is the PhD dissertation \cite{gachechiladze2019quantum}.

In this work one proposes some variations on the results of \cite{guhne2014entanglement} by studying SLOCC classification of four-qubit hypergraph states and violation of local realism of thoses states by considering Mermin's inequalities. In particular one shows that the SLOCC classes that can be obtained from hypergraph quantum states are very specific if one considers the geometry of Cayley's hyperdeterminant. We also show that the maximum violation of Mermin's inequalities is an efficient invariant to distinguish all the LU classes of four-qubit connected hypergraph states.

The paper is organized as follows. In Section \ref{hypergraph} one recalls the definition of hypergraph states and the four-qubit LU classification as it was obtained by \cite{guhne2014entanglement}
and \cite{chen2014locally}. In Section \ref{entanglement} one establishes which families of Vertraete et al.'s classification can be obtained from connected hypergraph states. Then we observe that only a specific part is concerned when one consider the four-qubit Hilbert space stratified by the singularities of the hyperdeterminant \cite{miyake_hyperdet}. In Section \ref{mermin} we determine numerically maximum violation of Mermin's inequalities for all connected hypergraph states and discuss the efficiency of this maximum amount of violation to distinguish the different LU-classes. Finally in Section \ref{ibm} one implements some of thoses inequalities on the IBM Quantum Experience.

\section{Hypergraph states}\label{hypergraph}
Let us recall the basic notions on hypergraph states that will be needed for this study. A good reference to start with is \cite{rossi2013quantum}.

\begin{figure}[!h]
 \includegraphics[width=10cm]{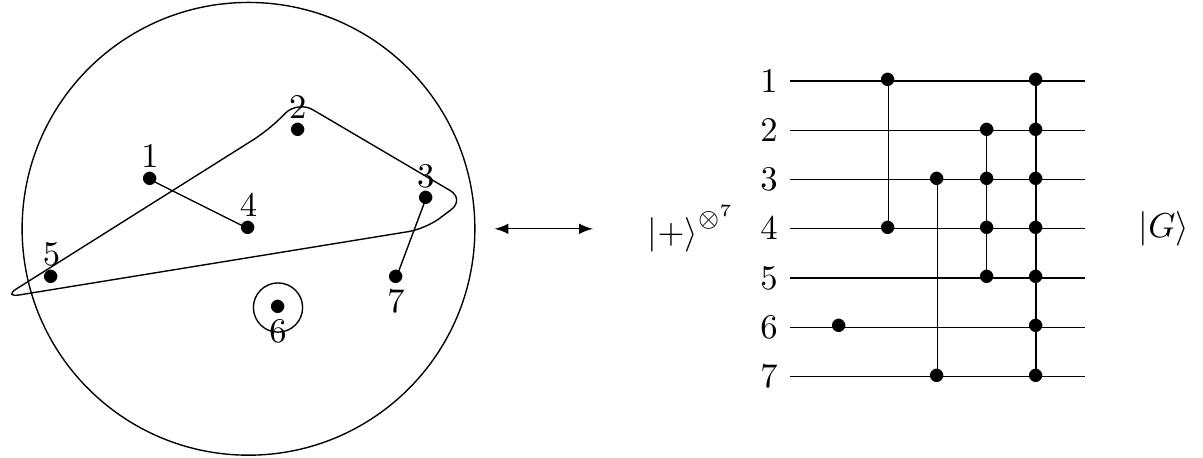}
 \caption{Example of hypergraph states. Left: An example of hypergraph with $V=\{1,2,3,4,5,6,7\}$ and $E=\{\{6\},\{1,4\},\{3,7\},\{2,3,4,5\},\{1,2,3,4,5,6,7\}\}$. Right: The implementing circuit generating the hypergraph state corresponding to the hypergraph on the left, with the initial state being $\ket{+}^{\otimes 7}$.}\label{fig:exhypergraph}
\end{figure}

A hypergraph is given by $G=(V,E)$ where $V=\{1,\dots,n\}$ is a set of vertices and $E\subset \mathcal{P}(V)$ is a set of hyperedges, i.e. a set of subsets of $V$. An example of such a hypergraph is given in Figure \ref{fig:exhypergraph} (left). To any hypergraph with $n=|V|$ one can associate a unique $n$-qubit hypergraph state by the following procedure: 
\begin{itemize} \item Consider the quantum state $\ket{+}^{\otimes n}=\frac{1}{\sqrt{N}}\sum_{i=0} ^{N-1} \ket{i}$ where $N=2^n$ and $\ket{i}$ is the decimal representation of the basis state $\ket{a_{n-1}a_{n-2}\dots a_1a_0}$ with $a_l\in \{0,1\}$, for $l=0,\dots,n-1$, and $i=a_{n-1}2^{n-1}+a_{n-2}2^{n-2}+\dots+a_12+a_0$.
\item For each hyperedge $e\in E$, such that $e=\{j_1,\dots,j_k\}$ apply a $C_eZ$-gate i.e. a control-$Z$ gate with control qubit $\{j_1,\dots,j_{k-1}\}$ and target qubit $\{j_k\}$ (by symmetry of the $C_eZ$ gate any qubit can be considered as the target qubit).
 \end{itemize}
 
This construction can be shortened by the following definition of a hypergraph state.
\begin{definition}
 Let $G=(V,E)$ a hypergraph and let us denote by $C_eZ$ the control-$Z$ gate associated to a hyperedege $e\in E\subset \mathcal{P}(V)$. Then the hypergraph state $\ket{G}$ is given by
 \begin{equation}
  \ket{G}=\prod_{e\in E} C_e Z\ket{+}^{\otimes n}.
 \end{equation}
\end{definition}

Graph states are examples of hypergraph states where hyperegdes are only of size two (edges), i.e. $e=\{j_1,j_2\}$. Like graph states, hypergraph states can be also defined within the stabilizer formalism, i.e. by describing  the abelian group of operators which stabilizes $\ket{G}$. 

\begin{figure}[!h]
 \begin{center}
  \includegraphics[width=10cm]{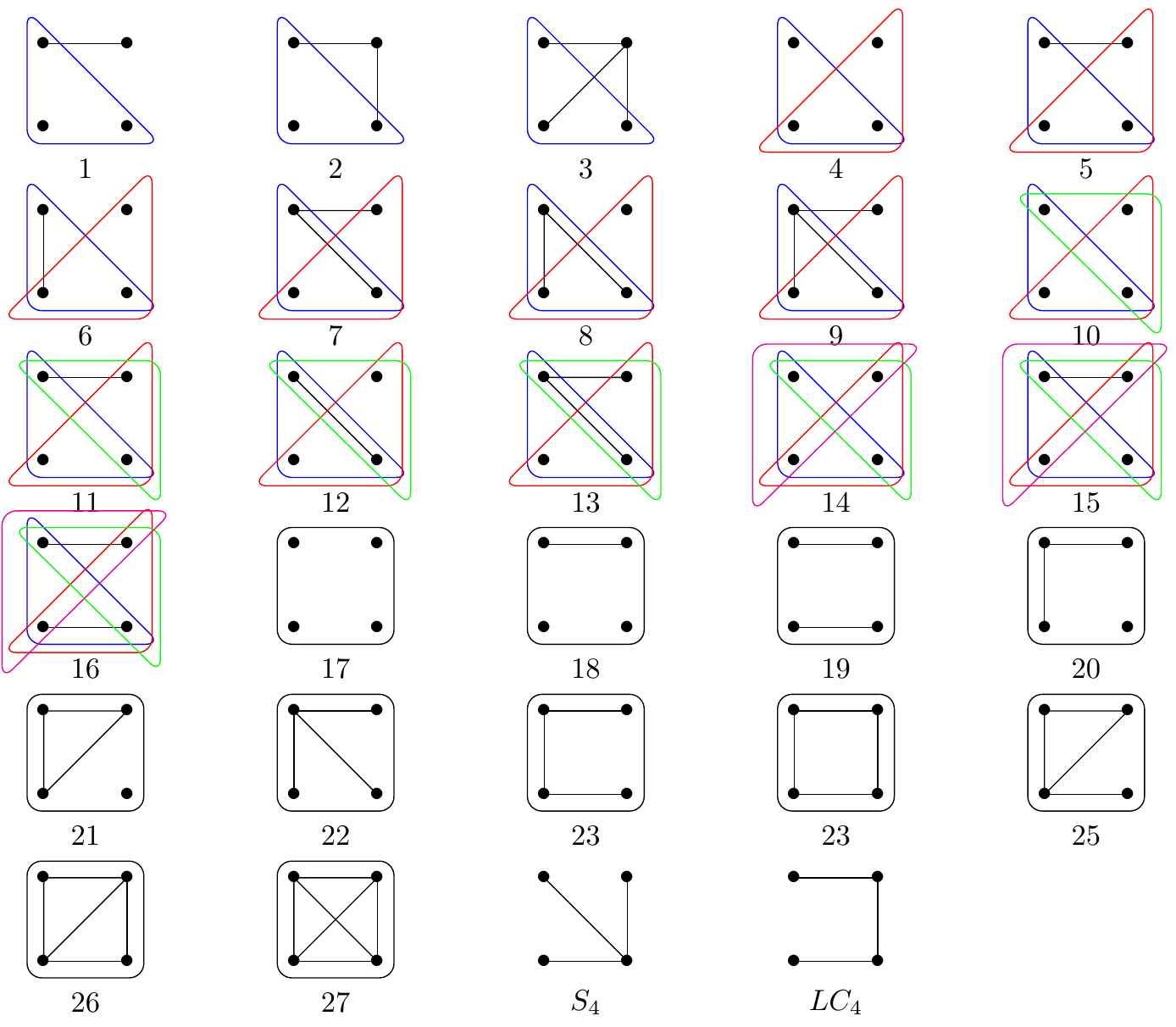}
 \end{center}
\caption{LU-classification of the connected four-qubit hypergraph states as established in \cite{guhne2014entanglement} and \cite{chen2014locally}. Hypergaphs 1--27 correspond to connected four-qubit hypergraph states with at least one hyperedge of size $3$. The graph states $S_4$ (star) and $LC_4$ (linear cluster) represent the only two classes of connected four-qubit graph states.}\label{fig:hyperclass}
\end{figure}

The number of possible hypergraph states for a given $n$ is $2^{2^n}$, i.e. $65,536$ hypergraph states in the four-qubit case. It can be first reduced by hypergraph isomorphism or equivalently permutation of qubits. In \cite{guhne2014entanglement,chen2014locally} the classification, up to local unitary transformation (LU) and permutation of qubits, for four-qubit connected hypergraph states was obtained. This classification contains $27$ classes involving at least one hyperedge of size $3$ and two classes corresponding to the connected four-qubit graph states. We reproduce the $27+2$ classes in Figure \ref{fig:hyperclass}.

\section{The four-qubit hypergraph SLOCC entanglement classes and Cayley's hyperdeterminant}\label{entanglement}
It was Miyake's original idea \cite{miyake_hyperdet} to use the singular locus of the hypersurface defined by Cayley's hyperdeterminant, denoted by $\text{HDet}$, to stratify the four-qubit Hilbert space. Recall that Cayley's hyperdeterminant is a special four-qubit SLOCC-invariant \cite{luque2003polynomial} also known as the defining equation of the dual variety $X^\vee$ of  $X=Seg(\PP^1\times\PP^1\times\PP^1\times\PP^1)\subset \PP(\CC^2\otimes\CC^2\otimes\CC^2\otimes\CC^2)$ the variety of separable states seen in the projectivization of the four-qubit Hilbert space \cite{holweck2012geometric}. For $\ket{\phi} \in \CC^2\otimes \CC^2\otimes \CC^2\otimes \CC^2$ let us denote by $H_\phi$ the projective linear space of states orthogonal to $\ket{\phi}$, i.e. $H_\phi=\PP(\{\ket{\psi} \in \CC^2\otimes \CC^2\otimes \CC^2\otimes \CC^2, \langle \phi,\psi\rangle=0\})$. Then the definition of HDet as equation of the dual of $X$ is equivalent to \cite{weyman1996,GKZ}: 
\begin{equation}
 \text{HDet}(\phi)\neq 0 \Leftrightarrow X\cap H_{\ket{\phi}} \text{ is a smooth hypersurface of } X
\end{equation}

More precisely one can also establish that 
 \begin{equation}\label{sing}
  \phi \text{ is a smooth point of HDet}=0 \text{ iff } X\cap H_\phi \text{ has a unique singularity of type } A_1.
 \end{equation}

One says that a hypersurface has a $A_1$ singularity at ${\bf x} \in \CC^n$ iff $f({\bf x})=0$, $\partial f_i ({\bf x})=0$ and the Hessian $(\partial ^2 _{ij} f({\bf x}))_{ij}$ is of full rank.

According to Eq. (\ref{sing}), a state $\ket{\phi}$ is a singular point of HDet$=0$ iff either $X\cap H_\phi$ has more than one singularity of type $A_1$ or has a unique singularity with higher degeneracy (starting with a degenerate Hessian). This leads to the definition of node and cusp component of $X^\vee$ as introduced in \cite{weyman1996}. We reformulate this definition in the context of the present study.

\begin{definition}
 Let $\ket{\phi} \in \CC^2\otimes\CC^2\times\CC^2\otimes \CC^2$ such that HDet$(\phi)=0$ and $\phi$ is not a smooth point of $X^\vee$. Then
 \begin{itemize}
  \item $\phi\in X^\vee _{\text{node}}$ iff $X\cap H_\phi$ has at least two singularities.
  \item $\phi\in X^\vee _{\text{cusp}}$ iff $X\cap H_\phi$ has a least one singularity  with degenerate Hessian.
 \end{itemize}
\end{definition}

Node and cusps components are also used in  the paper of Miyake \cite{miyake_hyperdet} as these components are SLOCC invariants. In fact one can show more precisely that the type of the singular hyperplane section is SLOCC invariant: In \cite{holweck2014singularity} using the classification of simple singularities of hypersurfaces, the singular type of each family of Verstraete's classification was computed, leading to a finer grained description of the singular locus of $X^\vee$ in terms of the types and corresponding family.

\begin{figure}[!h]
 \includegraphics[width=6cm]{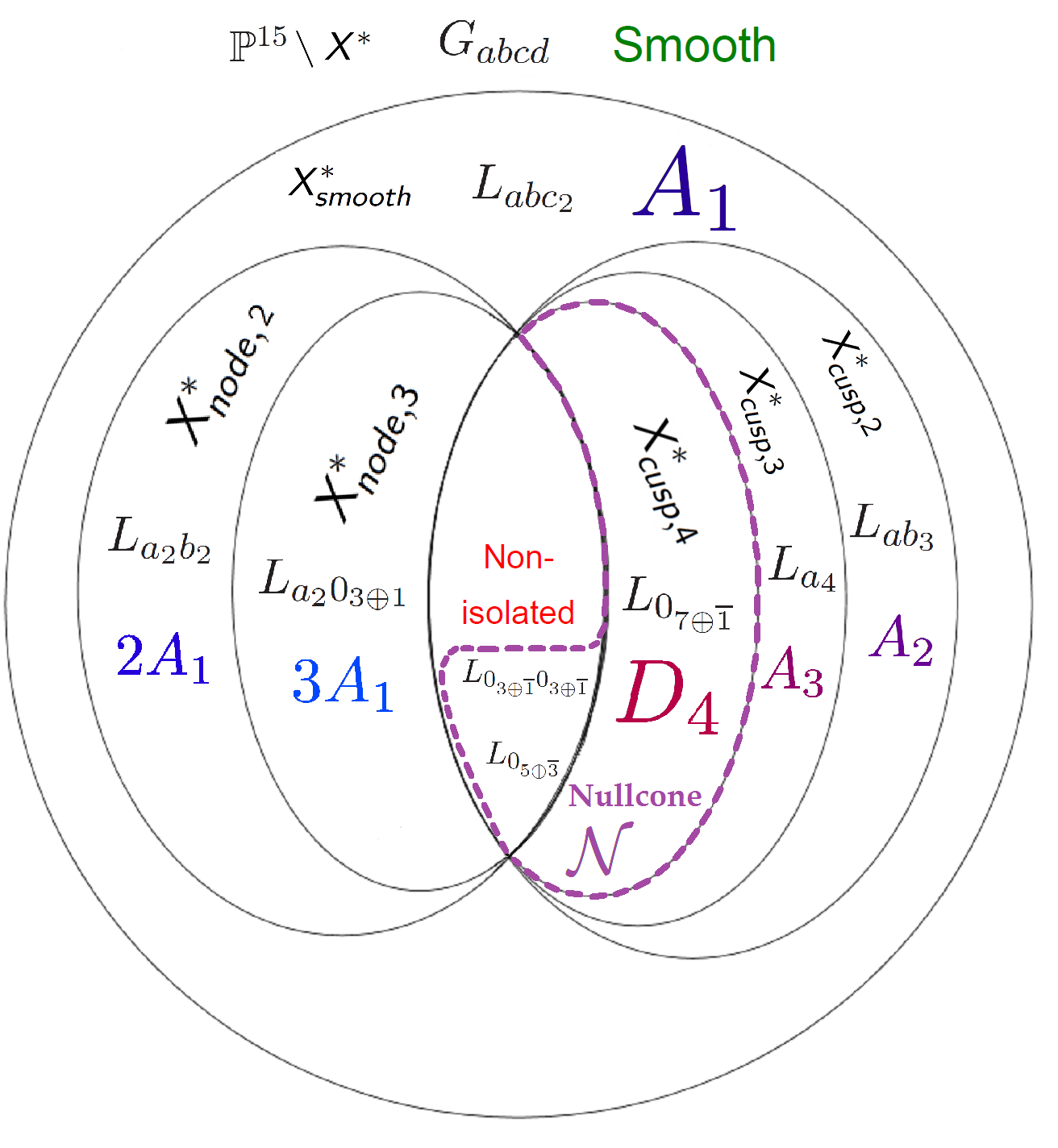}
 \caption{Stratification of the four-qubit Hilbert space in terms of the singular locus of  HDet: For $a,b,c,d$ generic, the family $G_{abcd}$ does not belong to $X^\vee$ and the corresponding hyperplane defines a smooth hyperplane section of $X$. For generic parameters $a,b,c$, the family $L_{abc_2}$ correspond to smooth points of $X^\vee$ i.e. hyperplane sections with a unique $A_1$ singularity. The families $L_{a_2b_2}$ and $L_{a_30\oplus1}$ belong to the node component and the families $L_{ab_3}$, $L_{a_4}$, $L_{0_{7\oplus \overline{1}}}$ are in the cusp component. All other families are in the intesection of the cusp and node sets. The nullcone is the subset of states that annihilate all SLOCC invariant polynomials. The notations are those of \cite{holweck2014singularity}.}\label{4onion}
\end{figure}

Figure \ref{4onion} encapsulates picturaly the finding of \cite{holweck2014singularity} and will be enough to explain the results of this section.

Back to the question of the SLOCC entanglement classes of connected four-qubit hypergraph states, we may ask the following question: which families of Verstraete's classification are reached by the hypergaph states of Figure \ref{fig:hyperclass} or equivalently which stratas defined by the singularities of HDet are obtained.

\begin{proposition}\label{prop:sing}
 Let $\ket{G} \in \CC^2\otimes\CC^2\otimes \CC^2\otimes \CC^2$ be a  connected four qubit graph or hypergraph state. Then either HDet$(G)\neq 0$ or $G\in X^\vee _{\text{node}}$ or $G$ is SLOCC equivalent to $ L_{0_{7\oplus \overline{1}}}$ . Moreover one can associate to each $27$ ($+2$) LU-hypergraph classes their corresponding SLOCC classes as described in Table \ref{SLOCCclass}.
 \end{proposition}
 \proof One applies the algorithm provided in \cite[Section V.]{holweck2017entanglement} to identify the SLOCC entanglement class of a given hypergraph state $\ket{G}$. This algorithm is based on invariants and covariants to identify the Verstraete's normal form of a given state. In terms of geometry the algorithm identifies which stratas with respect to $X^\vee$ and its singular locus is reached by a given state. For instance HDet$(G)\neq 0$ directly implies $\ket{G}\sim_{\text{SLOCC}}\ket{G_{abcd}}$. To double check our finding one also calculated the singular type of the hyperplane section $H_G$ for each connected hypergraph state $\ket{G}$ following \cite{holweck2014singularity}. Let us recall the principle of this calculation.
 
 The variety of separable states can be parametrized as follow by the Segre embedding \cite{holweck2014entanglement},
 \begin{equation}
  \resizebox{.9\hsize}{!}{$
 Seg:\left\{ \begin{array}{ccc}
    \PP^1\times\PP^1\times\PP^1\times \PP^1 & \to & \PP^{15}\\
    ([w _0:w _1],[x_0: x_1 ],[y_0 :y_1 ],[z_0:z_1 ]) & \mapsto & [w_0x_0y_0z_0:w_0x_0y_0z_1:w_0x_0y_1z_0:\dots:w_1x_1y_1z_1]
  \end{array}\right.$}
\end{equation}

A hyperplane $H_\phi$ defined by a state $\ket{\phi}=\sum_{i,j,k,l\in \{0,1\}} a_{ijkl} \ket{ijkl}$ will provide a hyperplane section of $X=Seg(\PP^1\times\PP^1\times\PP^1\times \PP^1)$ caracterized by the following equation:
\begin{equation}
 \sum_{i,j,k,l\in \{0,1\}} \overline{a}_{ijkl}w_ix_jy_kz_l=0
\end{equation}
The corresponding hypersurface of $\PP^1\times\PP^1\times\PP^1\times \PP^1$ can be seen as a hypersurface of $\CC\times\CC\times\CC\times \CC=\CC^4$ once we specify a chart. There are $16$ charts to consider. Let us detail two examples:

   We consider (in decimal notation) the hyperpgraph state $24$ of Figure \ref{fig:hyperclass}, i.e. \begin{equation}\ket{G_{24}}=\frac{1}{4}(\ket{0}+\ket{1}+\ket{2}-\ket{3}+\ket{4}+\ket{5}-\ket{6}+\ket{7}+\ket{8}-\ket{9}+\ket{10}+\ket{11}-\ket{12}+\ket{13}+\ket{14}-\ket{15}).\end{equation} The hyperplane section corresponding to $X\cap H_{G_{24}}$ is given in homogeneous coordinates by 
  \begin{equation} 
 \begin{array}{l}
  w_0x_0y_0z_0+ w_0x_0y_0z_1+w_0x_0y_1z_0-w_0x_0y_1z_1+w_0x_1y_0z_0+w_0x_1y_0z_1-w_0x_1y_1z_0\\+w_0x_1y_1z_1 
+w_1x_0y_0z_0-w_1x_0y_0z_1+w_1x_0y_1z_0+w_1x_0y_1z_1+w_1x_1y_0z_0+w_1x_1y_0z_1\\+w_1x_1y_1z_0-w_1x_1y_1z_1  =0.
\end{array}
\end{equation}
  In the chart $w_0=y_0=y_0=z_0=1$ this equation boils down to 
  \begin{equation}\label{eq:g24}
  \begin{array}{l}
  1+ z_1+y_1-y_1z_1+x_1+x_1z_1-x_1y_1+x_1y_1z_1 
+w_1-w_1z_1\\+w_1y_1+w_1y_1z_1+w_1x_1+w_1x_1z_1+w_1x_1y_1-w_1x_1y_1z_1  =0.
\end{array}
\end{equation}
  One can check that the hypersurface of $\CC^4$ defined by Eq. (\ref{eq:g24}) has no singularity and the same calculation in the $15$ other charts lead to the same conclusion. One concludes that $X\cap H_{G_{24}}$ is smooth which is also confirmed by the nonzero value of HDet$(\ket{G_{24}})$.

  Let us now consider the hypergraph state number $7$ of Figure \ref{fig:hyperclass}
  \begin{equation}\ket{G_7}=\dfrac{1}{4}(\ket{0}+\ket{1}+\ket{2}+\ket{3}+\ket{4}-\ket{5}+\ket{6}+\ket{7}+\ket{8}-\ket{9}+\ket{10}-\ket{11}+\ket{12}+\ket{13}-\ket{14}+\ket{15}).\end{equation}

 In the chart $w_0=x_0=y_0=z_0=1$ the equation defining the hypersurface is:
 \begin{equation}\label{eq:g7}
  \begin{array}{l}
  1+ z_1+y_1+y_1z_1+x_1-x_1z_1+x_1y_1+x_1y_1z_1 
+w_1-w_1z_1\\+w_1y_1-w_1y_1z_1+w_1x_1+w_1x_1z_1-w_1x_1y_1+w_1x_1y_1z_1  =0.
\end{array}
\end{equation}
This hypersurface has three isolated singularity of type $A_1$ at $(w_1,x_1,y_1,z_1)=(1+\sqrt{2},0,-1,1+\sqrt{2})$, $(w_1,x_1,y_1,z_1)=(1-\sqrt{2},0,-1,1-\sqrt{2})$,  and $(w_1,x_1,y_1,z_1)=(1,1,1,1)$. One can perform the same calculation on all the $15$ other charts and one gets that the hyperplane section $X\cap H_{G_7}$ has $4$ singularities of type $A_1$ (some singular points appear in serveral charts).

The third column of Table \ref{SLOCCclass} is obtained by performing the same calculation for all $29$ four-qubit connected hypergraph states and the second column is obtained by using the invariant/covariant algorithm of \cite[Section V.]{holweck2017entanglement}. All calculations are available at \url{https://quantcert.github.io/Mermin-hypergraph-states}. $\Box$
 \begin{table}[!h]
 \begin{tabular}{|c|c|c|c|c|}
 \hline
  Hypergraph class & Verstraete's family& Singularities& $\mu$ &$\tilde{\mu}$\\
  \hline
  1 & $L_{abc_2}$ ($a=b$, $c=0$)& $5A_1$ &$1.81129
$&3.28077\\
  2 & $L_{0_{7\oplus \overline{1}}}$ &$D_4$ &$1.26888
$&1.61650\\
  3 & $G_{abcd}$ ($d=0$) &Smooth &$1.76777
$&3.12500\\
  4 & $L_{abc_2}$ ($a=b$, $c=0$) &$5A_1$ &$1.93185
$
&3.73205\\
  $5$ &   $G_{abcd}$ ($c=d$, $d=0$) & $8A_1$& $1.83051$
&3.35078\\
$6$&	$Gr_7$& nonisolated &$1.22474$&1.50000\\
$7$ &$L_{a_{2}b_{2}}$&$4A1$ &$1.50000$
&2.28571\\
$8$ &$L_{a_{2}b_{2}}$ ($a=0$)& $4A_1$ &$1.50672
$&2.32137\\
$9$& $G_{abcd}$ ($d=0$)&  Smooth & $2.06066$&4.24632\\
$10$ &$L_{a_{2}0_{3\oplus 1}}$& $3A_1$ &$1.63359$&2.72222\\
$11$ & $L_{a_{2}b_{2}}$ & $4A_1$ & $1.89276
$&3.58437\\
$12$ &$L_{0_{7\oplus \overline{1}}}$ & $D_4$ & $1.35062
$&1.83211\\
$13$ &$L_{a_{2}b_{2}}$& $4A_1$ &$1.37175
$&1.88558\\
$14$ &$G_{abcd}$ ($d=0$)& Smooth  & $2.42329$&5.87234\\
$15$ &$G_{abcd}$ ($d=0$)&  Smooth &$1.55479
$&2.45225\\
$16$ &$G_{abcd}$ ($d=0$)& Smooth & $1.55430
$&2.53125\\
$17$ &$G_{abcd}$ ($a=b=c=0$)&   $6A_1$ &$1.43329$&2.07172\\
$18$ &$L_{abc_2}$ ($a=b,c=0$)&  $5A_1$&$1.31950
$
 &1.74308\\
$19$ &$G_{abcd}$ ($c=d,d=0$)& $4A_1$ &$1.84265
$&3.39919\\
$20$ &$L_{abc_{2}}$ ($a=0$)& $4A_1$ &$1.72283
$&2.96867\\
$21$ &$L_{a_{2}0_{3\oplus 1}}$& $3A_1$ &$1.70188
$&2.89678\\
$22$ &$L_{a_{2}0_{3\oplus 1}}$& $3A_1$ &$2.31759
$
&5.37105\\
$23$ &$L_{abc_{2}}$ ($c=0$) &$3A_1$&$1.59171
$ &1.66691\\
$24$ &$G_{abcd}$ ($d=0$)&  Smooth & $1.71310
$
&2.93497\\
$25$ &$L_{a_{2}b_{2}}$& $5A_1$ &$1.38608$&1.92164\\
$26$ &$G_{abcd}$ & Smooth &$1.49500
$&2.24320\\
$27$ &$G_{abcd}$&  Smooth &$2.21580$ &4.91327\\
\hline
$S_4$ &$G_{abcd}$ &$6A_1$  &$2.82843$
&8\\
$LC_4$&	$G_{abcd}$& $4A_1$ &$1.41421$
&2\\
  \hline
 \end{tabular}
 \caption{SLOCC families of the $27$ LU-classes of four-qubit connected hypergraph states and the two  connected four-qubit graph states $S_4$ and $LC_4$. The calculation of the singular type of the corresponding hyperplane sections confirms the fact that only points of $\PP(\CC^2\otimes\CC^2\otimes\CC^2\otimes\CC^2)\setminus X^\vee$,  the node component of $X^\vee$ and the orbit $L_{0_{7\oplus \overline{1}}}$ are reached by hyperpgraph states. The column ``Singularities'' describes the singularity of the corresponding hyperplane section. $A_1$ singularities are Morse points (singularities with Hessian matrix of full rank) and the $D_4$ singularity is an isolated singular point with Hessian matrix of corank $2$ and Milnor number equals $4$ (see \cite{holweck2014singularity} for more details on the calculation of the singular type). Columns $\mu$ and $\tilde{\mu}$ provide evaluations of Mermin's polynomials, see Sec. \ref{mermin}.}\label{SLOCCclass}
\end{table}
\begin{remark}
 It can be noticed that the number of $A_1$ singularities calculated for each hypergraph state is not the same as the one associated in Figure \ref{4onion} to the corresponding Verstraete's family. For instance the states of the family $L_{abc_2}$ correspond to smooth points of $X^\vee$ and therefore should have only one singular point of type $A_1$ while one obtains $5$ for the first four-qubit hypergraph state. This is not a contradiction in the sense that the results of Figure \ref{4onion} obtained in  \cite{holweck2014singularity} correspond to generic choice of parameters. The number of $A_1$ singularities calculated can be therefore higher than the one predicted by the family for specific choice of parameters.
\end{remark}

One may wonder if this pattern regarding the fact that hypergraph states belong essentially to the node component of $X^\vee$ remains for larger number of qubits. There is an analogue of Cayley's hyperdeterminant for $n\geq 5$. It is the hyperdeterminant of format $(2,2,\dots,2)$ and geometrically it corresponds the defining equation  of the dual variety of $X=\PP^1\times\dots\times \PP^1$. For $n\geq 5$, there is no known expression of the hyperdeterminant HDet$_{2,...,2}$ but the caracterization of smooth, node or cusp points in terms of the singularities of the corresponding hyperplane section is the same. As there is no classification of hypergraph states for $n\geq 5$, we restrict ourselves to one type of hypergraph states.

\begin{definition}
 One says that a $n$-hypergraph states $\ket{G}$, with $G=(V,E)$, is the $k$-uniform hypergraph states of size $n$ iff $E$ only contains  hyperedge of size $k$ and contains all of them.
\end{definition}

We have conducted similar calculations as in Proposition \ref{prop:sing} for the singularities of the hyperplane sections of $k$-uniform hypergraph states for $n=5,6,7$. Our results are summerized in Table \ref{tab:kuniformsing}. The Table indicates that some pattern may be found. For instance one can easly prove that the singular type of the hyperplane sections for $2$-uniform and $n$-uniform $n$ hypergraph states will always be composed of non isolated singularities. Indeed the singular type of the hyperplane section is SLOCC invariant and both $2$-uniform and $n$-uniform $n$-qubit hypergraph states are SLOCC equivalent to $\ket{GHZ_n}$. With $x_0^{(i)}, x_1^{(i)}$ being the coordinates of the $i$-th  copy of $\CC^2$ in $\mathcal{H}=(\CC^2)^{\otimes n}$, the homogeneous hyperplane section corresponding to $X\cap H_{GHZ_n}$ is given by:
\begin{equation}
 x^{(1)}_0x_0^{(2)}\dots x_0^{(n)}+x^{(1)}_1x_1^{(2)}\dots x_1^{(n)}=0.
\end{equation}
For $n\geq 5$ the corresponding hypersurface will always have nonisolated singularities. This is already the case if one considers the chart $x^{(1)}_1=x^{(2)} _1=x^{(3)} _1=1$ and $x^{(4)}_0=\dots=x^{(n)}_0=1$.

More insteresting is the outcome of our calculation for $3$-uniform hypergraph states. For $n=5,6,7$ the hyperplane section are always smooth meaning that the corresponding state $\ket{G}$ does not belong to the dual variety (HDet$(G)\neq 0$). More work would be needed to prove this result in full generality.  

\begin{table}[!h]
 \begin{tabular}{|c|ccc|}
 \hline
    &$n=5$ & $n=6$ & $n=7$ \\
 $k=2$ & Non-isolated & Non-isolated & Non-isolated \\
 $k=3$ & Smooth & Smooth & Smooth \\
 $k=4$ & $A_1$ & $A_m$, $m>1$ & $A_m$, $m>1$ \\
 $k=5$ & Non-isolated & ?? & ??\\
 $k=6$ &    & Non-isolated & ??\\
 $k=7$ &    &              & Non-isolated\\
 \hline
 \end{tabular}
\caption{Type of singular section of $k$-uniform hypergraph states for $n=5,6,7$. $A_m$-singularities, $m>1$, correspond to singularities of hypersurfaces with Hessian of corank one. The corresponding hyperplanes belong to the cusp component of the hypersurface defined by HDet. The question marks indicate that our Maple calculation was unconclusive.}\label{tab:kuniformsing}
\end{table}


\section{Mermin's inequalities and hypergaph states}\label{mermin}
Non-local properties of hypergaph states have been first studied in \cite{guhne2014entanglement,gachechiladze2016extreme}. In particular in \cite{guhne2014entanglement} it was shown how the stabilizer formalism can be used to derive for hypergraph states new Bell-like inequalities and in \cite{gachechiladze2016extreme} it was proved, based on those inequalities, that $k$-uniform hypergraph states maximally violate those inequalities.

Here instead of using the stabilizer formalism to design Bell-like inequalities we will use  Mermin's polynomials. To each hypergraph state we associate the maximum value that  Mermin's polynomials can achieve when we optimize over all possible choice of measurements. That will be enough in order to  show that all connected four-qubit hypergraph states exhibit non-local properties. This maximum value, denoted by $\mu$ see Table \ref{SLOCCclass}, is a LU-invariant that can be employed to distinguish the different LU-classes.

Let us recall the definition of Mermin's polynomials and how it can be used to prove that a given state is non-local \cite{collins2002bell}.

\begin{definition}\label{def:mermin}
 Let $\{a_1,a_1',a_2,a_2',\dots\}$ be a family of two-qubit observables. Mermin's polynomials are defined inductively as:
 \begin{itemize}
  \item $M_1=a_1$
  \item $M_n=\dfrac{1}{2}M_{n-1}(a_n+a_n')+\dfrac{1}{2}M_{n-1}'(a_n-a_n')$
 \end{itemize}
 with $M_{n-1}'$ obtained from $M_n$ by interchanging primed and nonprimed observables.
\end{definition}

Under the hypothesis of local realism (LR) the maximum value that can be reached is $1$ while it is $2^{\frac{n-1}{2}}$ under the assumption of Quantum Mechanics. This leads to the Mermin's inequalities:
\begin{equation}
 \langle M_n\rangle _{LR}\leq 1 \hspace{1cm} \langle M_n\rangle _{QM}\leq 2^{\frac{n-1}{2}}.
\end{equation}
Let us now consider Mermin's polynomials as polynomials depending on parameters, i.e. depending on the choice of the observables $\{a_1,a_1',a_2,a_2',\dots\}$. For each $i$ let us denote by $\alpha_i, \beta_i, \gamma_i$ the three real parameters such that $a_i=\alpha_iX+\beta_i Y+\gamma_i Z$ and $\alpha_i^2+\beta_i ^2+\gamma_i^2=1$ where $X,Y,Z$ are the usual Pauli matrices.

For a given state $\ket{\psi}\in (\CC^2)^{\otimes n}$ one can numerically compute the following LU-invariant:
\begin{equation}\label{eq:mu}
 \mu(\psi)=\text{Max}_{\alpha_i^2+\beta_i^2+\gamma_i^2=1,\alpha_i'^2+\beta_i'^2+\gamma_i'^2=1}\langle \psi |M_n(\alpha_1,\beta_1,\dots,\gamma_n')|\psi\rangle
\end{equation}

We provide in Table \ref{SLOCCclass} the different values of $\mu$ computed by a random walk algorithm\footnote{A similar calulation was recently done in \cite{de2020mermin} to measure non-local behavior of states generated by Grover's algorithm. Both calculations of \cite{de2020mermin} and the present paper are available at \url{https://quantcert.github.io/}.}.

Let us make some observations:
\begin{enumerate}
 \item The non-locality of all four-qubit connected hypergraph states can be detected by Mermin's polynomials as we found $\mu>1$ for all of them. 
 \item As a LU-invariant $\mu$ allows us to distinguish the $29$ classes of the Table. A consequence is that a classification algorithm of the four-qubit connected hypergraph states could be implemented based on this evaluation.
 \item The highest values of $\mu$ are obtained for $k$-uniform four-qubit hypergraph states for $k=2$ and $k=3$.
\end{enumerate}
Non-locality of $3$-uniform and $4$-uniform hypergraph states were studied in \cite{gachechiladze2016extreme} for asymptotic values of $n$ by considering specific Bell like inequalities built on the stabilizer formalism. If we restrict to small value of $n$ one can numerically estimate $\mu$ for all $k$-uniform hypergraph states. Table \ref{tab:kuniform} gives the results we obtained up to $n=9$.
\begin{table}[!h]
 \begin{tabular}{|c|ccccc|}
 \hline
 & $n=5$ & $n=6$ & $n=7$ & $n=8$ &$n=9$\\
$k=2$ &$4$ &$5.65685$ &$8$ & $11.31370$& $16$\\
$k=3$ &$2.45751$ &$2.85947$ &$4.34159$ &$6.24393$ &$8.2368$  \\
$k=4$ & $2.02319$&$3.29038$ &$4.51349$ & $4.92526$ &$6.0113$\\
$k=5$ &$1.29200$ &$3.20848$ &$5.93197$ & $8.97846$&$11.6284$ \\
$k=6$ & &$1.14326$ &$2.44886$& $3.69746$&$5.7151$ \\
$k=7$ & & &$1.00307$ & $3.17162$ &$6.9736$\\
$k=8$ & & & &$0.87610$ & $2.4187$ \\
$k=9$ & & & & &$0.7430$ \\
\hline
 \end{tabular}
\caption{Numerical estimation of $\mu$ for $k$-uniform hypergraph states with $n\in \{5,\dots,10\}$ and $2\leq k\leq 10$.}\label{tab:kuniform}
\end{table}

An other LU-invariant of interest, based on Mermin's polynomials, is the following quantity:
\begin{equation}
 \tilde{\mu}(\psi)=\text{Max}(\langle \psi |M_n|\psi\rangle^2+\langle \psi |M_n'|\psi\rangle^2)
\end{equation}
where like in Eq. (\ref{eq:mu}) the Max is obtained over all coefficients $\alpha_i,\dots,\gamma_i'$. This quantity was studied in \cite{yu2003classifying,endrejat2005characterization} where it was shown to be useful to detect some specific type of entanglement making an interesting connection between non-locality measure (Mermin's polynomials evaluation) and global entanglement. More precisely the following sufficient conditions regarding entanglement of a four-qubit states $\ket{\psi}$, can be obtained from a more general result of \cite{yu2003classifying}:
\begin{itemize}
 \item $\ket{\psi}$ is $2$-entangled, i.e. is a product of two $2$-entangled qubits or the product of one $2$-entangled pair  with two single qubits. Then $\tilde{\mu}\leq 2$.
 \item $\ket{\psi}$ is $3$-entangled, i.e. is the product of a genuine entangled three-qubit state with a single qubit. Then $\tilde{\mu}(\psi)\leq 4$.
 \item $\ket{\psi}$ is $4$-entangled, i.e. is genuine four-qubit entangled state. Then $\tilde{\mu}\leq 8$.
\end{itemize}
One notices that $\tilde{\mu}$ also distinguish the $29$ classes of Table \ref{SLOCCclass} if one considers the results within a $10^{-2}$ approximation. Moreover the calculation of $\tilde{\mu}$ detects the genuine four-entanglement for the states in the set \begin{equation}\{9,14,22,27,S_4\},\end{equation} and shows that the states 
\begin{equation}\{1,3,4,5,7,8,11,15,16,17,19,20,21,24,26\}\end{equation} are at least $3$-entangled.

\section{Implementation on the IBM Quantum Computer}\label{ibm}
One interesting aspect of Mermin's inequalities is that it can be  evaluated on a quantum computer. Violation of Mermin's inequalities for $\ket{GHZ_n}$-like states was first established by Alsina {\em et al.} in \cite{alsina2016experimental} for $n=3,4,5$. Similar calculation for $W$-state was done in \cite{swain2019experimental}. In \cite{cervera2019quantum} the generation by quantum circuits of maximally entangled states is discussed and an improvement of the violation of Mermin's inequalities for $\ket{GHZ_5}$ is obtained.

The IBM Quantum Experience\footnote{\url{https://quantum-computing.ibm.com/}} proposes to the users a graphical interface which allows  to perform quantum computation in the circuit formalism. To compute experimentally the values of Mermin's polynomials for four-qubit hypergraph states, one needs to generate each hypergraph state and then perform a measurement for each monomial involved in the calculation of the given Mermin's polynomial. In the four-qubit case a Mermin's polynomial involves $16$ monomials and therefore one would need to produce $16\times 29 =464$ different circuits if one wants to check with the IBM quantum computer all calculations of $\mu$ in Table \ref{SLOCCclass}.

In order to gain in efficiency we have used the open-source software Qiskit which allows us to program the needed calculation.  Our commented sources are available at \url{https://quantcert.github.io/Mermin-hypergraph-states} and can be used by the reader. There are essentially two programs, one generating the circuit of a  given hypergraph state from the description of the hyperedges and an  other program performing and collecting, from the circuit of a given hypergraph state,  the $16$ measurements necessary to evaluate $\mu$.

\subsection{Circuit of hypergaph states}
\begin{figure}[!h]
 \includegraphics[width=8cm]{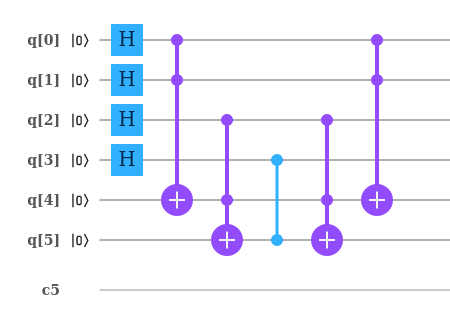}
 \caption{Circuit generating $\ket{G_{17}}$ the hypergraph state $17$ with only one hyperedge of size $4$. The first four wires correspond to the generation of $\ket{G_{17}}$ while $q[4], q[5]$ are auxiliary qubits. One reads from the circuit that a $Z$ gate is applied to $q[3]$ iff the control auxiliary qubit $q[5]=1$. But $q[5]=1$ iff $q[2]$ and $q[4]$ equal $1$ (Toffoli). Finally $q[4]=1$ iff $q[0]=q[1]=1$. Thus the $Z$ gate is applied to $q[3]$ iff $q[0]=q[1]=q[2]=1$ i.e. it corresponds to a $c-c-c-Z$ gate or a four-qubit hyperedge. The last two Toffoli gates are necessary to disentangle the auxiliary qubits $q[4]$ and $q[5]$ from the main part of the circuit.}\label{circuit:hyper4}
\end{figure}

Circuits to generate graph states with the IBM Quantum Experience can be implemented straightforwardly as the gates $H$ and $c-Z$ are available. Similarly $c-c-Z$ gate can be implemented with two Hadamard gates and one Toffoli. However there is no multiple $c-c-\dots-c-Z$ gates predefined. Such multiple controlled $Z$ gate can be obtained using several Toffoli gates and auxiliary qubits as shown in Figure \ref{circuit:hyper4}.

In Qiskit pseudo-code the generation of hypergraph states can be expressed by the Algorithm \ref{algo:hypergraph} in Appendix \ref{sec:appendix}. The generation of multiple control $Z$ gates corresponds to the loop within the condition $|e|>3$ (hyperedge of size at least $4$) and involves the use of auxilary qubits. Note that the last {\bf while} loop is necessary to disentangle the main circuit from the auxiliary qubits.


\subsection{Measuring Mermin's polynomials with Qiskit}
The next step consists in realizing the different measurements to evaluate each monomial of Mermin's polynomial. Let us consider a monomial $a_1a_2a_3a_4$ corresponding to four directions $\vec{v}_1,\vec{v}_2,\vec{v}_3, \vec{v}_4$ on the Bloch sphere, i.e. $a_i=\alpha_i X+\beta_i Y+\gamma_i Z$ with $\alpha_i,\beta_i,\gamma_i$ reals such that $\alpha_i^2+\beta_i^2+\gamma_i^2=1$ and 
$\vec{v}_i=(\alpha_i,\beta_i,\gamma_i)$. In order to compute the expectation of $a_1a_2a_3a_4$ one needs to measure the ith-qubit in the $\vec{v}_i$ direction. However the IBM Quantum Experience only allows measurement in the $Z$-basis, i.e. in the 
$\vec{v}=(0,0,1)$ direction. To deal with it one needs to find the unitary matrix corresponding to the change of basis from the direction $\vec{v}_i$ to $\vec{v}$. This can be achieved with the following gate implemented on the IBM Quantum Experience.
\begin{equation}
 U_3(\theta,\phi,\lambda)=
 \begin{pmatrix}
                           \cos(\theta/2) & e^{-i\lambda}\sin(\theta/2)\\
                           e^{i\phi}\sin(\theta/2) & e^{i(\phi+\lambda)}\cos(\theta/2)
\end{pmatrix}
\end{equation}
\begin{lemma}
 Let us consider a direction $\vec{v}_i$ on the Bloch sphere given in spherical 
 coordinates by $\vec{v}_i=(\cos(\theta_i)\sin(\phi_i),\sin(\theta_i)\sin(\phi_i),\cos(\phi_i))$. Then measuring a single qubit $\ket{\psi}$ in the direction $\vec{v}_i$ is equivalent to measuring the qubit $U_3(\theta_i,\pi,-\phi_i-\pi)\ket{\psi}$ in the $Z$-basis.
\end{lemma}

\proof The proof is straightforward. The change of basis from the $Z$-basis to the one defined by the $\vec{v}_i$ direction is given by 
$P=\begin{pmatrix}
   \cos(\theta_i/2) & -\sin(\theta_i/2)\\
   e^{i\phi_i}\sin(\theta_i/2) & e^{-i\phi_i}\cos(\theta_i/2)                                                                                                                    \end{pmatrix}.$ Then it is clear that $U_3(\theta,\pi,-\phi-\pi)=P^\dagger$, leading to the result. $\Box$\\
   
\begin{figure}[!h]
 \centering
 \includegraphics[width=12cm]{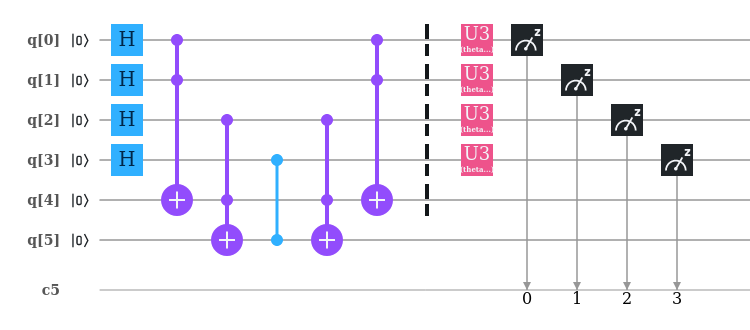}
 \caption{Measurement of the $a_1a_2a_3a_4$ monomial. For each operator $a_i$ the direction $\vec{v}_i$ is calculated by the numerical evaluation of $\mu$. Then the $U_3$ matrices are placed on the circuit to realize the measurements in the $\vec{v}_i$ directions.}\label{fig:measurement}
 \end{figure}

The choice of the directions $\vec{v}_i$ are dicted by the optimization process corresponding to the numerical evaluation of $\mu$ (Eq. (\ref{eq:mu}) and Table \ref{SLOCCclass}). For each monomial given by a collection of four directions, one places the corresponding $U_3$ gates on the circuit to implement the correct measurement (Figure \ref{fig:measurement}).
This second step of the algorithm is summarized in Algorithm \ref{algo:mermin}.

\subsection{Results}
The IBM Quantum Experience allowed us to run our calculation on a simulator or to send our calculation to one of the IBM quantum machine.

\subsubsection{Simulator}
One tested on the simulator the evaluation of $\mu$ on the $29$ four-qubit hypergraph states of Figure \ref{fig:hyperclass}. Up to a $10^{-2}$ precision the IBM Quantum simulator provided the same result as the numerical evaluation obtained in Table \ref{SLOCCclass}. Because the codes are written using Qiskit one can also check on the IBM Quantum simulator the evaluation of $\mu$ for $k$-uniform hypergraph states. For instance we were able to recover on the simulator with a $10^{-2}$ precision all results of Table \ref{tab:kuniform}.

\subsubsection{Quantum machine}
When we delegate to the IBM Quantum Machine the evaluation of each monomials, the results are not as good as with the simulator. Before executing a given circuit, there is the {\em transpilation} step which translates the circuit to an 
equivalent calculation on the quantum machine. This transpilation process is needed to take into consideration the specific architecture of the quantum machines used in the IBM Quantum Experience. For instance each Toffoli gate is {\em transpiled} to a circuit involving $5$ $c-X$ gates and several rotations. The transpiled version of the hypergraph state of Figure \ref{circuit:hyper4} involves $32$ $c-X$ gates and $18$ rotations gates. It shows how in practice the optimization of CNOT-circuits is an important problem for quantum computing \cite{bataille2020quantum}.
\begin{figure}[!h]
 \includegraphics[width=17cm]{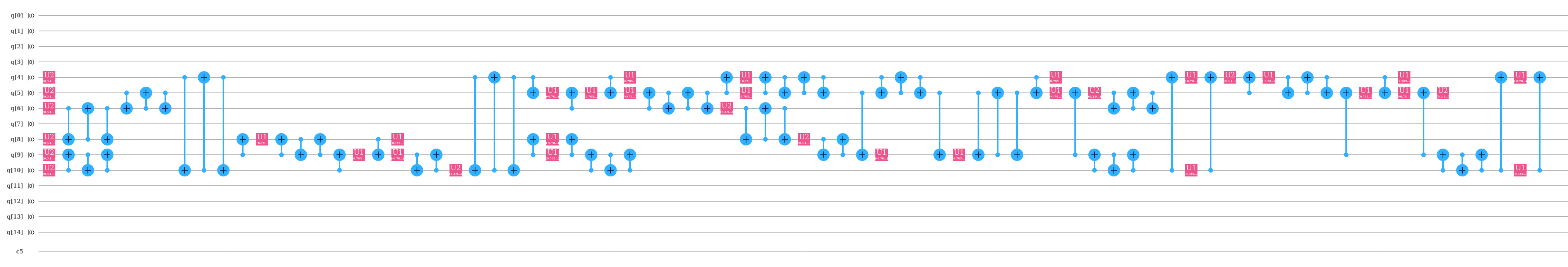}
 \caption{Transpiled version of the circuit represented in Figure \ref{circuit:hyper4} when implemented on the quantum machine {\bf ibmq\_16\_melbourne}.}
\end{figure}

\begin{figure}[!h]
 \includegraphics[width=3cm]{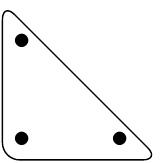}
 \includegraphics[width=8cm]{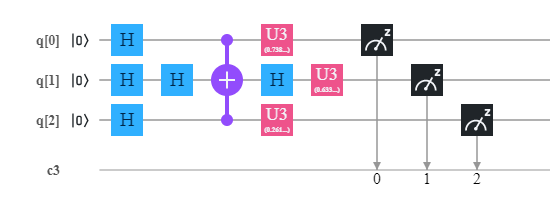}
 \caption{The first example of hypergraph state that is not a graph state and the circuit representing the evaluation of one monomial.}\label{fig:3hyper}
\end{figure}

In fact, despite the accuracy of the measures obtained with the simulator, we were not able to obtain violation of Mermin's inequalities for any hypergraph states of Figure \ref{fig:hyperclass}. 
The only example of violation of Mermin's polynomial we were able to obtain with a hypergraph is for the $3$-qubit case (Figure \ref{fig:3hyper}). In this case the Mermin's polynomial is given by
\begin{equation}
 M_3=\frac{1}{2}(ABC'+AB'C+A'BC-A'B'C'),
\end{equation}
with 
\begin{equation}\begin{array}{ccc}
 A=0.58 X+0.44Y-0.68Z &A'=0.37X-0.83Y-0.41Z\\
 B=-0.58 X-0.44Y+0.68Z &A'=-0.37X+0.83Y+0.41Z\\
 C=0.58 X+0.44Y-0.68Z &C'=0.37X-0.83Y-0.41Z\\
\end{array}\end{equation}

The evaluation of $M_3$ on the hypergraph state $\ket{G}=c-c-Z\ket{+}^{\otimes 3}$ gives \begin{equation}\label{eq:munum} \mu(\ket{G})\approx 1.52.\end{equation} The evaluation of the four monomials with the IBM Quantum Experience produces the following results:
\begin{equation}
\begin{array}{cc}
 \langle A'BC\rangle=0.52 & \langle AB'C\rangle=0.62\\
 \langle ABC'\rangle=0.64 & \langle A'B'C'\rangle=-0.48
\end{array}
\end{equation}
Which provides the following value of $\mu$:
\begin{equation}\label{eq:muexp}
 \mu_{exp}\approx 1.13.
\end{equation}
As expected the experimental evaluation of $\mu$, Eq. (\ref{eq:muexp}), is not as accurate when compared to the numerical evaluation (Eq. (\ref{eq:munum})). But the experimental value violates Mermin's inequalities as $\mu_{exp}\approx 1.13>1$ providing, to the best of our knowledge, a first example of experimental violation of local realism with a hypergraph state on the IBM Quantum Experience.

\section{Conclusion}

In this paper we studied entanglement and nonlocality of four-qubit hypergraph states. One discussed the SLOCC entanglement classes that are achieved by four-qubit connected hypergraph states and examine the singular type associated to those states with respect to the stratification of the four-qubit Hilbert space induced by Cayley $2\times 2\times 2\times 2$ hyperdeterminant. One also considered evaluation of nonlocality by calculating numerically the maximum value obtained by Mermin's polynomials when evaluated on a four-qubit connected hypergraph state. We intended to implement those calculations on the IBM Quantum Experience. As mentioned in the introduction, this paper can be considered as a variation of the original work of \cite{guhne2014entanglement} which tackles the LU-classification as well as asymptotical behavior of non-locality for some type of hypergraph states. Our work can also be seen in connection with \cite{bataille2019quantum} where SLOCC entanglement classes of states generated by circuits only made of SWAP and $c-Z$ gates is investigated. In \cite{bataille2019quantum}, HDet as well as the algorithm  used in Proposition \ref{prop:sing} for the SLOCC classification, are also considered.

Finding direct connection between algebraic invariants, entanglement properties and measure of nonlocality is not a straightforward task. Even if connections exist there is no one-to-one correspondence between their quantitative evaluation. For instance one can observe that the four-qubit hypergraph states that does not vanish Cayley's hyperdeterminant (Sec. \ref{entanglement}), HDet, have relatively high value of the nonlocality measure $\tilde{\mu}$ (Sec. \ref{mermin}) but at this stage one can barely provide more conclusion connecting the two types of calculations provided in this work. 
However we believe that it is worth to keep considering algebraic invariants, like HDet, as valuable tools to study quantum properties that  could be evaluated on a quantum machine. In this respect the recent work  of \cite{perez2020measuring}  opens interesting perspectives. In a future work one would like to implement similar calculation to evaluate HDet on the IBM Quantum Experience like we did for Mermin's polynomials. 
\section{Acknowledgements}
This work was supported by the R\'egion Bourgogne Franche-Comt\'e, project PHYFA (contract 20174-06235), the French Investissements d'Avenir programme, project ISITE-BFC (contract ANR-15-IDEX-03) and the EUR-EIPHI Graduate School (Grant No. 17-EURE-0002). The authors acknowledge the use of the IBM Quantum Experience. The views expressed are those of the authors and do not reflect the official policy or position of IBM or the IBM Quantum Experience team. The authors would also like to thank the developers of the open-source framework Qiskit. We  thank our colleague Henri de Boutray for our exchanges regarding the Qiskit implementations.
\bibliographystyle{plain}

\bibliography{biblio}

\appendix
\section{Algorithms}\label{sec:appendix}
In this Appendix, one provides pseudo-code versions of our Qiskit codes to create hypergraph states and to evaluate on a quantum state a  Mermin polynomial given by the parameters of the set of observables $\{a_1,a_1',\dots, a_n,a_n'\}$ (See Definition \ref{def:mermin}). The codes and their corresponding documentations are available at \url{https://quantcert.github.io/Mermin-hypergraph-states}.
\begin{algorithm}[!h]
 \KwData{A hypergraph states given by $(V,E)$ with longest hyperedege of size $k$}
 \KwResult{The quantum circuit generating $\ket{G}$}
 $n=|V|$\;
  circuit=QuantumCircuit($n$,$n$)\;
\For{$i=0$ to $n-1$}{circuit=circuit.h($i$)}
\For{$e\in E$}{\If{$|e|=2$}{circuit.cz(h($0$),h($1$))}
\If{$|e|=3$}{circuit.h($2$)\;
    circuit.toffoli(h($0$),h($1$),h($2$))\;
   circuit.h($2$)\;}
\If{$|e|>3$}{cpt=0\;
  max=$|e|-1$\;
  qubit\_aux=$n$\;
  circuit.toffoli(e(cpt),e(cpt+1),qubit\_aux)\;
  cpt=$+2$\;
 \While{cpt$<$max}{circuit.toffoli(e(cpt),qubit\_aux,qubit\_aux+1)\;
 qubit\_aux=qubit\_aux+1\;
 cpt=cpt+1\;
 circuit.cz(qubit\_aux,e(cpt))\;
 qubit\_aux=qubit\_aux-1\;
 cpt=cpt-1\;}
 \While{max$>$2}{circuit.toffoli(qubit\_aux-1,e(cpt),qubit\_aux)\;
 qubit\_aux=qubit\_aux-1\;
 cpt=cpt-1\;
 max=max-1\;}
 circuit.toffoli(e(cpt),e(cpt-1),qubit\_aux)\;
 }}
 \caption{Qiskit pseudo-code for generating hypergraph states. The hyperedge of size $2$ and $3$ can be obtained from the implemented $c-Z$ and $c-c-Z$ gates (obtained from the Toffoli gate). For hyperedges of size at least $4$ one needs to introduce auxiliary qubits. The last {\bf while} loop is necessary to disentangle the auxiliary qubits from the main circuit.}\label{algo:hypergraph}
\end{algorithm}

\begin{algorithm}[!h]
 \KwData{A circuit generating a graph state $\ket{G}$}
 \KwResult{Evaluation of $\mu$ on the IBM Quantum Experience simulator/machine}
  Compute by random walk algorithm the parameters of the observables $a_1,a_1'\dots,a_n,a_n'$ that maximizes $\mu$\;
  Transform the parameters as set of directions $v_1, v_1',\dots,v_n,v_n'$\;
  Generate from Defintion \ref{def:mermin} a vector $M$ of size $2^n$ encoding the  coefficients of the monomials of $M_n$\; \# To each index $i$ correspond a possible monomial of $M_n$ with the following rule: Let us consider the binary expension of $i$ on $n$ bits, $i=b_{n-1}b_{n-2}\dots b_0$. Then the monomial corresponding to the index $i$ will be $c_1\dots c_n$ where $c_j=a_j$ if $b_{n-j}=0$ and $c_j=a_j'$ if $b_{n-j}=0$\;
  res=0\;
 \For{$i=0$ to $2^n-1$}{
 {\If{$M[i]\neq 0$}{Place the gates $U_3$ according to the description of the monomial of index $i$ and the corresponding directions\;
 Measure on the IBM Quantum Simultor/Machine\;
 eval=Collect the data to evaluate the mean value of the monomial of index $i$\;
 res=res+$M[i]\times$eval:\;}
 }}
 \caption{Pseudo-code corresponding to the evaluation of $M_n$ on a hypergraph states $\ket{G}$.}\label{algo:mermin}
\end{algorithm}

\end{document}